\documentclass[sigconf]{acmart}

\renewcommand\footnotetextcopyrightpermission[1]{} 
\usepackage{amsmath}
\usepackage{amsthm}
\usepackage{amsfonts}
\usepackage{graphicx}
\usepackage{subfig}
\usepackage{listings}
\usepackage{siunitx}

\def\eg{\textit{e.g.}}
\def\ie{\textit{i.e.}}

\def\dpl{\texttt{diffprivlib}}
\def\Dpl{\texttt{Diffprivlib}}
\def\ver{0.1.1}

\keywords{Differential privacy, Python, open source, machine learning, data analysis.}

\begin{document}

\title{\Dpl: The IBM Differential Privacy Library}
\subtitle{A general purpose, open source Python library for differential privacy}

	\author{Naoise Holohan}
	\orcid{0000-0003-2222-9394}
	\affiliation{%
	\institution{IBM Research -- Ireland}
	}
	\email{naoise@ibm.com}
	  \author{Stefano Braghin}
	  \affiliation{%
	\institution{IBM Research -- Ireland}
	}
	\email{stefanob@ie.ibm.com}

	 \author{P\'{o}l Mac Aonghusa}
	  \affiliation{%
	\institution{IBM Research -- Ireland}
	}
	\email{aonghusa@ie.ibm.com}

	\author{Killian Levacher}
	 \affiliation{%
	\institution{IBM Research -- Ireland}
	}
	\email{killian.levacher@ibm.com}

\begin{abstract}
Since its conception in 2006, differential privacy has emerged as the de-facto standard in data privacy, owing to its robust mathematical guarantees, generalised applicability and rich body of literature.
Over the years, researchers have studied differential privacy and its applicability to an ever-widening field of topics.
Mechanisms have been created to optimise the process of achieving differential privacy, for various data types and scenarios.
Until this work however, all previous work on differential privacy has been conducted on a ad-hoc basis, without a single, unifying codebase to implement results.

In this work, we present the IBM Differential Privacy Library, a general purpose, open source library for investigating, experimenting and developing differential privacy applications in the Python programming language.
The library includes a host of mechanisms, the building blocks of differential privacy, alongside a number of applications to machine learning and other data analytics tasks.
Simplicity and accessibility has been prioritised in developing the library, making it suitable to a wide audience of users, from those using the library for their first investigations in data privacy, to the privacy experts looking to contribute their own models and mechanisms for others to use.
\end{abstract}

\maketitle

\section{Introduction}
For more than a decade, differential privacy has been a focal-point for the research and development of new methods for data privacy.
Buoyed by the general applicability of differential privacy, and the rigorous mathematical privacy guarantees it possesses, researchers have contributed to a comprehensive body of literature proposing new applications for differential privacy in a wide array of fields.
In the early days of differential privacy, simple applications included privacy-preserving histograms~\cite{Dwo06}, while in more recent times differential privacy has been applied to deep learning~\cite{ACG16}.
In between, seemingly countless adaptations of machine learning models have been presented that satisfy differential privacy, presenting a genuine opportunity to perform privacy-preserving exploration or analysis of data.

However, these applications of differential privacy currently exist in a fragmented development environment, encompassing different programming languages, diverse coding styles and incomplete, unmaintained, model-by-model, repositories that makes the implementation of differential privacy a difficult task, even for experts in the field\footnote{Examples of valuable contributions to differential privacy include \url{https://github.com/uber/sql-differential-privacy}, \url{https://github.com/brubinstein/diffpriv} and \url{https://gitlab.com/dp-stats/dp-stats/}, but all use different programming languages.}.
With the work presented herein, we are bringing together a number of differential privacy applications and fundamentals that will be accessible and useful to all levels of expertise and knowledge on data privacy.

The IBM Differential Privacy Library, also known as \dpl, is written in Python 3, a popular programming language for machine learning and data analysis.
\Dpl\ leverages the functionality and familiarity of the NumPy~\cite{WCV11} and Scikit-learn~\cite{PVG11} packages, meaning functions and models are instantly recognisable, with default parameters ensuring accessibility for all.
Released under the MIT Open Source license, \dpl\ is free to use and modify, and contributions of its users are welcomed to help expand the functionality and features of the library.

\Dpl\ provides a large collection of mechanisms, the fundamental building blocks of differential privacy that handle the addition of noise.
These mechanisms are used under-the-hood in the machine learning models and other tools to satisfy differential privacy, allowing the complex task of achieving differential privacy to be hidden from view.
Like Scikit-learn, machine learning models can be trained in just two lines of code with \dpl: one \texttt{import} statement, and one line to fit the model (\eg, \texttt{LogisticRegression().fit(X, y)}).

In this paper we provide a brief overview of the library (Section~\ref{sc:overview}), give details on the library's main modules (Section~\ref{sc:modules}) and provide a worked example (Section~\ref{sc:example}) to showcase its functionality.

\section{Overview}\label{sc:overview}
\Dpl\ is written in Python, and requires Python 3.4 to run.
The library is not compatible with Python 2, which is due to reach end-of-life in 2020.
Python was selected due to its accessibility, wealth of machine learning capabilities, and active user community.
The library is written to comply with the PEP8 coding standard~\cite{PEP8}, as much as is practicable.

\Dpl\ can be installed using the \texttt{pip} command:
\begin{lstlisting}
>>> pip install diffprivlib
\end{lstlisting}

A major pillar of \dpl\ is for it to be accessible to researchers and programmers of all privacy expertise, from those investigating differential privacy for the first time, to the more advanced users looking to develop their own applications.
To this end, we have integrated \dpl\ with the popular machine learning library Scikit-learn, allowing users to run machine learning instances in a near-identical way to how they do so already.
Similarly, we have mirrored and leveraged the familiarity and functionality of NumPy in developing easy-to-use differential privacy tools.

The library consists of three main modules:
\begin{itemize}
\item \texttt{mechanisms} -- a collection of differential privacy mechanisms, the building blocks for developing differential privacy applications;
\item \texttt{models} -- a collection of differentially private machine learning models;
\item \texttt{tools} -- a collection of tools and utilities for simple data analytics with differential privacy.
\end{itemize}

\Dpl\ is hosted on GitHub\footnote{\url{https://github.com/IBM/differential-privacy-library}}, and cloning the library from there provides users with additional resources.
A \texttt{tests/} directory includes unit tests for library functionality, and can be run using \texttt{pytest} from the root directory after installation.
A collection of Jupyter notebook examples is also included in the \texttt{notebooks/} directory that showcase the functionality of the library.
Documentation is hosted on Read~the~Docs\footnote{\url{https://diffprivlib.readthedocs.io/}}.

\section{Library Contents}\label{sc:modules}

\subsection{Mechanisms}
The mechanisms that have been implemented in Version~\ver\ of \dpl\ are listed in Table~\ref{tbl:mechanisms}.
These mechanisms have been built primarily for their inclusion in more complex applications and models, but can be executed and experimented with in isolation, as shown in Listing~\ref{lst:mechanism}.

For simplicity, similar mechanisms are grouped together in the same sub-module (\eg, all Laplace-like mechanisms are contained in \texttt{mechanisms.laplace}), but all can be imported directly from\linebreak \texttt{mechanisms}.

Mechanisms are interacted with using function-specific methods, reducing the need for code and documentation duplication.
For example, each mechanism has a \texttt{set\_epsilon()} (and\slash or\linebreak \texttt{set\_epsilon\_delta()}) method to set the $\epsilon$ (and $\delta$) parameters of the mechanism, and many mechanisms have a\linebreak \texttt{set\_sensitivity()} method.
Similarly, each method has a\linebreak \texttt{randomise()} method, which takes an input value and returns a differentially private output value, assuming the mechanism has been correctly configured.
A mechanism's methods can be checked in the documentation (\eg, \texttt{help(Binary)}), or by using Python's \texttt{dir()} method.

\begin{lstlisting}[caption={Example usage of \texttt{mechanisms}},float,label={lst:mechanism}]
>>> <\textbf{from}> diffprivlib.mechanisms <\textbf{import}> Laplace

>>> laplace = Laplace()
>>> laplace.set_epsilon(0.5)
>>> laplace.set_sensitivity(1)
>>> laplace.randomise(3)
5.835104866820303
\end{lstlisting}

\begin{table}[t]
 \centering
 \begin{tabular}{|l|p{3.5cm}|} 
 \hline
 \textbf{Mechanism name} & \textbf{Notes} \\
 \hline\hline
 \texttt{DPMechanism} & Base class for all mechanisms\\
 \hline
 \texttt{TruncationAndFoldingMixin} & Mixin for truncating or folding numeric outputs of a mechanism\\
 \hline\hline
 \texttt{Binary} & \cite{HLM17} \\
 \hline
 \texttt{Exponential} & \cite{MT07} \\
 \hline
 \texttt{ExponentialHierarchical} & \texttt{Exponential}, with hierarchical utility function \\
 \hline
 \texttt{Gaussian} & \cite{DR14} \\
 \hline
 \texttt{GaussianAnalytic} & \cite{BW18} \\
 \hline
 \texttt{Geometric} & \cite{GRS12} \\
 \hline
 \texttt{GeometricTruncated} & \texttt{Geometric}, with post-processing\\
 \hline
 \texttt{GeometricFolded} & \texttt{Geometric}, with post-processing and support for half-integer bounds\\
 \hline
 \texttt{Laplace} & \cite{DMN06}, with support for $\delta > 0$ from \cite{HLM15}\\
 \hline
 \texttt{LaplaceTruncated} & \texttt{Laplace}, with post-processing\\
 \hline
 \texttt{LaplaceFolded} & \texttt{Laplace}, with post-processing\\
 \hline
 \texttt{LaplaceBoundedDomain} & \cite{HAB18}\\
 \hline
 \texttt{LaplaceBoundedNoise} & \cite{GDG18}\\
 \hline
 \texttt{Staircase} & \cite{GKO15}\\
 \hline
 \texttt{Uniform} & Special case of~\cite{GDG18} when $\epsilon=0$\\
 \hline
 \texttt{Vector} & \cite{CMS11}\\
 \hline
\end{tabular}
\caption{The list of mechanisms included in \dpl\texttt{.mechanisms}.}
\label{tbl:mechanisms}
\end{table}

\subsection{Machine Learning Models}
The models developed in \dpl\ have been engineered to mirror the behaviour and syntax of the privacy-agnostic versions present in Scikit-learn.
This allows for a simple one-step change to switch from a vanilla Scikit-learn machine learning model to one which implements differential privacy with \dpl.

For example, when implementing a Gaussian na\"ive Bayes classifier with differential privacy, a user need only replace the \texttt{import} statement from Scikit-learn with the corresponding import from \dpl\ and run their code and analysis in the same way.
\Dpl\ supports the same pre-processing pipelines that are present in Scikit-learn, and, in some cases, their use is encouraged to optimise noise-addition and model sensitivity.
The $\epsilon$ value is specified as a parameter when the model is initialised (\eg,\linebreak \texttt{GaussianNB(epsilon=0.1)}), otherwise a default of $\epsilon =1$ is used.

\begin{lstlisting}[caption={Example usage of \texttt{models}},float,label={lst:models}]
>>> <\textbf{from}> diffprivlib.models <\textbf{import}> GaussianNB

>>> nb = GaussianNB()
>>> X = [[-1, -1], [-2, -1], [-3, -2], [1, 1], [2, 1], [3, 2]]
>>> y = [1, 1, 1, 2, 2, 2]
>>> nb.fit(X, y)
PrivacyLeakWarning: Bounds have not been specified ...

>>> nb.predict([[-0.5, -2], [3,2]])
array([1, 2])
\end{lstlisting}

For more advanced users, the models can be parameterised with information about the data to ensure no privacy leakage (and the accompanying \texttt{PrivacyLeakWarning}) and optimal noise-addition.
This can include specifying the range of values for each feature (column) of the data, or the maximum norm of each sample in the data.
If these parameters are not specified in advance of training the model, the user is warned about a possible privacy leak.
However, the model continues to execute, with the parameters computed on the data.
For the example given in Listing~\ref{lst:models}, the warning could have been suppressed by initialising with \texttt{GaussianNB(bounds=[(-3, 3), (-2, 2)])}, as the values in each column of \texttt{X} fall within those bounds. 

The differentially private machine learning models that have been implemented in Version~\ver\ of \dpl\ are listed in Table~\ref{tbl:models}.

\begin{table}[t]
 \centering
 \begin{tabular}{|l|p{4.5cm}|} 
 \hline
 \textbf{Model} & \textbf{Notes} \\
 \hline\hline
 \multicolumn{2}{|c|}{Supervised learning}\\
 \hline
 \texttt{LogisticRegression} & Implements the logistic regression classifier in~\cite{CMS11} (using the \texttt{Vector} mechanism), with minor changes to allow for non-unity data norm and to allow integration with the corresponding classifier in SKLearn.\\
 \hline
 \texttt{GaussianNB} & Implementation of the Gaussian na\"{i}ve Bayes classifier presented in~\cite{VSB13} (using the \texttt{Laplace} and \texttt{LaplaceBoundedDomain} mechanisms) with minor amendments\footnote{The paper omitted splitting $\epsilon$ between the randomisation operations acting on the mean and variance of each feature, something we account for with a uniform split}\\
 \hline\hline
 \multicolumn{2}{|c|}{Unsupervised learning}\\
 \hline
 \texttt{KMeans} & Implementation of the $k$-means algorithm presented in~\cite{SCL16} (using the \texttt{GeometricFolded} and \texttt{LaplaceBoundedDomain} mechanisms)\\
 \hline
\end{tabular}
\caption{The list of machine learning models included in \dpl\texttt{.models}.}
\label{tbl:models}
\end{table}

\subsection{Tools}
The \texttt{tools} module of \dpl\ contains a number of basic tools and utilities that include differential privacy.
Presently, the module includes histogram functions to compute differentially private histograms on data, and statistical functions for computing the differentially private mean, variance and standard deviation of an array of numbers.
All of these tools mirror and leverage the functionality of the corresponding functions in NumPy, ensuring broad applicability and ease of use.

Further details are given in Table~\ref{tbl:tools}.

\begin{table}[t]
 \centering
 \begin{tabular}{|p{2cm}|p{5.5cm}|} 
 \hline
 \textbf{Tools} & \textbf{Notes} \\
 \hline\hline
 \texttt{histogram}, \texttt{histogram2d}, \texttt{histogramdd} & Histogram functions mirroring and leveraging the functionality of their NumPy counterparts, with differential privacy (using the \texttt{GeometricTruncated} mechanism)\\
 \hline
 \texttt{mean}, \texttt{var}, \texttt{std} & Simple statistical functions mirroring and leveraging their Numpy counterparts, with differential privacy (using the \texttt{Laplace} and \texttt{LaplaceBoundedDomain} mechanisms)\\
 \hline
\end{tabular}
\caption{The list of differential privacy tools included in \dpl\texttt{.tools}.}
\label{tbl:tools}
\end{table}

\subsection{Troubleshooting}
Two library-specific warnings that can be triggered by \dpl:
\begin{enumerate}
\item \texttt{PrivacyLeakWarning}: Triggered when a differential privacy application (in \texttt{models} or \texttt{tools}) has been incorrectly configured and consequently will not strictly satisfy differential privacy.
In such circumstances, the execution of the application continues, but the user is warned about the potential for additional privacy leakage (beyond that controlled by differential privacy).
Examples of a \linebreak\texttt{PrivacyLeakWarning} include: (i) not specifying the bounds\slash norm of the data when calling a model (requiring the model to compute the bounds\slash norm on the data, thereby leaking privacy); or (ii) where the bounds\slash norm supplied to the model do(es) not cover all the data, resulting in a potential violation of differential privacy.
\item \texttt{DiffprivlibCompatibilityWarning}: Triggered when a parameter, typically present in the parent of the function\slash class (\ie, in NumPy or Scikit-learn), is specified to a \dpl\ component which does not use the parameter.
The user is warned of this incompatibility to ensure full understanding of the results, and continues execution with the parameter amended\slash deleted as required by \dpl.
\end{enumerate}

\section{Worked Example}\label{sc:example}
We now present an example use case of the \dpl.
In this example we use the na\"ive Bayes classifier \texttt{GaussianNB},  implemented as a child class of Scikit-learn's \texttt{GaussianNB()}.
Na\"ive Bayes is a probabilistic classifier that learns the means and variances of each feature (assumed independent) for each label, allowing Bayes theorem to be applied to classify unseen examples.
Differential privacy is applied by adding noise to the means and variances that the model learns, ensuring a decoupling of the model from the data upon which it was trained.

The first step in using the model is to import the training and test data we wish to use.
For this simple example, we use the Iris flower dataset, and implement an 80\slash 20 train\slash test split,  using \texttt{sklearn}.

\begin{lstlisting}
>>> <\textbf{from}> sklearn <\textbf{import}> datasets
>>> <\textbf{from}> sklearn.model_selection <\textbf{import}> train_test_split

>>> dataset = datasets.load_iris()
>>> X_train, X_test, y_train, y_test = train_test_split(dataset.data, dataset.target, test_size=0.2)
\end{lstlisting}

We can now use \texttt{X\_train} and \texttt{y\_train} to train the differentially private model.
To do so, we import \texttt{GaussianNB()} from \dpl\ and call its \texttt{fit()} method.
The model will fall back on default parameters if none are specified at initialisation.

\begin{lstlisting}
>>> <\textbf{from}> diffprivlib.models <\textbf{import}> GaussianNB

>>> clf = dp.GaussianNB()
>>> clf.fit(X_train, y_train)
PrivacyLeakWarning: Bounds have not been specified ...
\end{lstlisting}

The \texttt{PrivacyLeakWarning} is triggered as a result of not specifying the bounds of the data on initialisation.
The model must therefore compute the bounds on the given data, resulting in additional privacy leakage.
To prevent this, the bounds should be determined independently of the data (\ie, using domain knowledge), and specifying them as a parameter at initialisation.

We can now classify unseen examples, knowing the model now satisfies differential privacy and protects the privacy of the individuals in the training dataset.
The resulting array is the predicted class of the corresponding row in \texttt{X\_test}.

\begin{lstlisting}
>>> clf.predict(X_test)
array([1, 0, 2, 1, 2, 1, 2, 1, 0, 0, 1, 2, 2, 0, 0, 0, 1, 1, 1, 1, 0, 2, 1, 1, 0, 0, 1, 0, 0, 1])
\end{lstlisting}

We can check the accuracy of the prediction using the corresponding \texttt{y\_test} array.

\begin{lstlisting}
>>> (clf.predict(X_test) == y_test).sum() / y_test.shape[0]
0.9333333333333333
\end{lstlisting}

\begin{figure}[t]
	\includegraphics{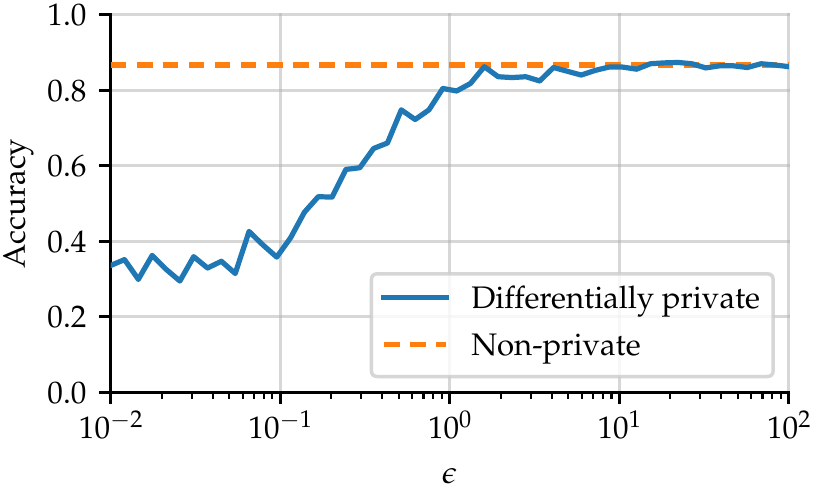}
  \caption{\label{fig:gaussiannb} Comparison of accuracy versus $\epsilon$ for a differentially private na\"ive Bayes classifier on the Iris dataset. For each $\epsilon$, the average accuracy over \num{30} simulations is shown.}
\end{figure}

We can loop through values of $\epsilon$ to plot the accuracy of the model for a range of privacy guarantees, as plotted in Figure~\ref{fig:gaussiannb}.
The small size of the Iris flower dataset (\num{150} samples) results in a high threshold for $\epsilon$ in order to maintain accuracy.
Similar analysis on the larger UCI Adult dataset (\num{48842} samples) shows significantly superior accuracy, reaching 95\% of the non-private accuracy at $\epsilon=0.05$.

\section{Conclusion}
In this paper we have presented the IBM Differential Privacy Library, a general purpose, open source Python library for the simulation, experimentation and development of applications and tools for differential privacy.
We have presented the functionality of Version~\ver\ of the library, but also its user-friendly development ethos, smooth integration with the popular NumPy and Scikit-learn packages and wider development potential which will be built upon in the forthcoming months and years.
We encourage readers to install the library (with \texttt{pip install diffprivlib}), and start running their own user-friendly, differentially private machine learning tasks.
As an open source project, we welcome and encourage contributions to develop and enhance the library in all their forms.

\section*{Acknowledgements}
A special thanks goes to our former colleagues Maria-Irina Nicolae and Spiros Antonatos for their instrumental help in getting \dpl\ off the ground, and to our colleague Joao Bettencourt-Silva for his help in testing the library.

\end{document}